\documentclass{article}%[fullpage,draft]{IEEEtran}
\usepackage{pstricks,pst-plot,psfig,fancybox, fullpage}

\makeatletter

\def\@abssec#1{\vspace{.05in} \parindent 0in {\bf #1. }\ignorespaces}
\def\abstract{\@abssec{Abstract}}

\def\keywords{\@abssec{Key words}}

\def\AMSMOS{\@abssec{AMS(MOS) subject classifications}}

\def\@begintheorem#1#2{\par\vskip .1 in\bgroup{\sc\bf #1\ #2. }\it\ignorespaces}
\def\@opargbegintheorem#1#2#3{\par\bgroup{\sc #1\ #2\ (#3). }\it\ignorespaces}
\def\@endtheorem{\egroup}

\newtheorem{lemma}{Lemma}
\newtheorem{theorem}{Theorem}
\newtheorem{definition}{Definition}
\newtheorem{corollary}{Corollary}

\makeatother

\def\nowidth#1{{\setbox0=\hbox{#1}\hspace*{-\wd0}\hbox to 0pt {#1}}}
\def\nospace#1{{\vbox to 0.0in{\hbox to 0.0in {#1}}}}
\def\fixed#1{{\vbox to 0.0in{\hbox to 0.0in{\tiny #1}}}}

\begin{document}
\title{Index Assignment for Multichannel Communication under Failure%Matrix Problem
} %!PN

\author{Tanya Y. Berger-Wolf\thanks{Department of Computer Science,
                               University of Illinois at Urbana-Champaign,
                               1304 W. Springfield Avenue,
                               Urbana, Illinois 61801, USA.
			Supported in part by an NSF Graduate Fellowship and
			by NSF grant CCR-95-30297.
                              {\tt tanyabw@uiuc.edu}},
Edward M. Reingold\thanks{Department of Computer Science,
                               University of Illinois at Urbana-Champaign,
                               1304 W. Springfield Avenue,
                               Urbana, Illinois 61801, USA.
   				Supported in part by NSF grant CCR-95-30297.
				{\tt reingold@cs.uiuc.edu}}}

\maketitle
\begin{abstract}
We consider the problem of multiple description scalar quantizers and
describing the achievable rate-distortion tuples in that setting. We
formulate it as a combinatorial optimization problem of arranging
numbers in a matrix to minimize the maximum difference between the
largest and the smallest number in any row or column. We develop a
technique for deriving lower bounds on the distortion at given channel
rates. The approach is constructive, thus allowing an algorithm that
gives a closely matching upper bound. For the case of two
communication channels with equal rates, the bounds coincide, thus
giving the precise lowest achievable distortion at fixed rates. The
bounds are within a small constant for higher number of channels. To
the best of our knowledge, this is the first result concerning systems
with more than two communication channels.

%We consider the problem of communication over multiple channels in which
%channels can fail and the information sent over those channels is lost. The
%goal is to minimize the difference between the received and the original
%information while having
%as little redundancy as possible. The information is encoded into an
%ordered tuple, each component of which is sent over a separate 
%channel.
%We analyze the problem of designing an encoding that minimizes the decoding 
%error in case of channel failure. We design and algorithm
%for the case of equal channel capacities and specific range of numbers and
%design a generic technique for proving lower bounds on the error in general.
\end{abstract}


%\begin{AMSMOS}
%94A05, 94A24, 94A34, 68Q25, 68R10
%\end{AMSMOS}

\input epsf

\section{Introduction.}
\begin{figure}
\begin{center}
\psset{unit=1.5}{
\begin{pspicture}(0,-2)(7,2)
\rput[br](0,0){\Large $x$}
\rput[tr](0,-.1){$\log m$}
\psline{->}(0,0.1)(0.4,0.1)\psframe(0.4,-.5)(1.6,.7)\rput(1,0.1){ENCODE}
\psline{->}(1.6,0.1)(2,0.1)\psline{-}(2,-1.9)(2,2.1)
\rput[r](1.9,2.1){channel $1$}\psline{-}(2,2.1)(2.5,2.1)\rput(2.65,2.1){$i_1$}
       \rput(3.5,2.2){$\log n_1$}\psline{->}(3,2.1)(4,2.1)
       \rput(4.4,2.1){$i_1$}\psline{-}(4.5,2.1)(5,2.1)
\rput[r](1.9,1.3){channel $2$}\psline{-}(2,1.3)(2.5,1.3)\rput(2.65,1.3){$i_2$}
       \rput(3.5,1.4){$\log n_2$}\psline{->}(3,1.3)(4,1.3)
       \rput(4.4,1.3){$i_2$}\psline{-}(4.5,1.3)(5,1.3)
\psline{-}(2,0.5)(2.5,0.5)\psline{->}(3,0.5)(4,0.5)\psline{-}(4.5,0.5)(5,0.5)
\psline{-}(2,-0.3)(2.5,-0.3)\rput(2.65,-0.3){$i_j$}
       \psline{->}(3,-0.3)(4,-0.3)\rput(3.5,-0.3){$\times$}
       \rput(4.4,-0.3){$*$}\psline{-}(4.5,-0.3)(5,-0.3)
\psline{-}(2,-1.1)(2.5,-1.1)\psline{->}(3,-1.1)(4,-1.1)\psline{-}(4.5,-1.1)(5,-
1.1)
\psdots[dotscale=.5](2.65,.7)(2.65,.4)(2.65,.1)(4.4,.7)(4.4,.4)
(4.4,.1)
\rput[r](1.9,-1.9){channel $k$}\psline{-}(2,-1.9)(2.5,-1.9)\rput(2.65,-1.9){$i_k$}
       \rput(3.5,-1.8){$\log n_k$}\psline{->}(3,-1.9)(4,-1.9)
       \rput(4.4,-1.9){$i_k$}\psline{-}(4.5,-1.9)(5,-1.9)
\psline{-}(5,-1.9)(5,2.1)\psline{->}(5,0.1)(5.4,0.1)
\psframe(5.4,-.5)(6.6,.7)\rput(6,0.1){DECODE}\psline{->}(6.6,0.1)(7,0.1)
\rput[bl](7.1,0){\Large $\tilde x$}
\end{pspicture}
}
\end{center}
\caption{Schematic setting of the multichannel problem.}
\label{multichannel}
\end{figure}
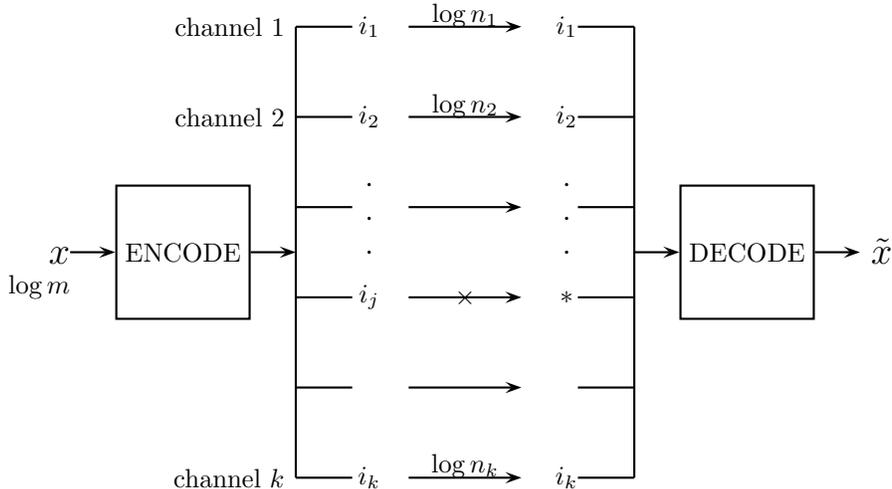

Consider sending information over $k$ independent unreliable channels.
We want to partition the source information into $k$ subsets so that
if all $k$ subsets are received, the original information can be
completely reconstructed, and if any of the channels fail, then the
error, defined as the absolute (rather than Hamming) difference
between the original information and the possible reconstruction of
it, is minimized. Figure \ref{multichannel} shows the general setting
of the problem.  A trivial solution would be to divide the source
information into $k$ equal blocks, sending each over a separate
channel; however, if any block fails to arrive, that part of the
information is lost completely. The error in this case could strongly
depend on which channel was lost, a feature we would like to
avoid. Alternatively, we could send $k$ complete copies, so that even
if only one of the channels succeeds, all the information is still
available; however, while the scheme is robust it utilizes the
resources poorly.  Our goal is to partition the information in a way
that allows us to recover as closely as possible the information
originally sent, the error depending on the number of channels lost,
regardless of which channels failed.

\subsection{Background}
The problem of designing codes for a {\em diversity-based}
(multichannel) communication system, that guarantee a minimum fidelity
at the user end, based on the number of channels succeeding in
transmitting information, is known as the {\em Multiple Descriptions
problem}. It was introduced by Gersho, Witsenhausen, Wolf, Wyner, Ziv,
and Ozarow at the 1979 IEEE Information Theory Workshop. It is a
generalization of the classical problem of source coding subject to a {\em
fidelity criterion} \cite{shannon}.

Initial progress on the Multiple Descriptions problem was made by
El-Gamal and Cover \cite{gamal}, who studied the the achievable rate
region for a memoryless and a single-letter fidelity criterion. Ozarow
\cite{ozarow} showed that the achievable region derived in \cite{gamal}
is the rate-distortion region for the special case of a memoryless
Gaussian source and a square-error distortion criterion.  Zhang and
Berger \cite{zhang} and Witsenhausen, Wolf, Wyner, and Ziv \cite{WW},
\cite{WWZ} explored whether the achievable rate region is the
rate-distortion region for other types of information sources.

The first constructive results for two channels with equal rates were
presented by Vaishampayan \cite{vaish}, \cite{vaish2}. In \cite{vaish}
Vaishampayan designs Multiple Descriptions Scalar Quantizers (MSDQs)
with good asymptotic properties. We show, however, that this solution
is not optimal.

An MSDQ is a {\em scalar quantizer} (mapping of the source to a finite
integer point set) that is designed to work in a diversity-based
communication system. The problem of designing an MSDQ consists of two
main components: constructing an {\em index assignment}, which is a
mapping of an integer source to a tuple to be transmitted over the
channels, and optimizing the structure of the quantizer for that
assignment.  This paper focuses on the index assignment problem. We
present a general technique for designing index assignments for any
number of channels with arbitrary rates. We give upper and lower
bounds for the information distortion for fixed channel rates. In case
of two channels transmitting at equal rates, the bounds coincide, thus
giving an optimal algorithm for the index assignment problem. In the
case of three or more equal-rate channels, the bounds are within a
multiplicative constant.

Real applications of the index assignment problem arise in video and
speech communication over packet-switched networks, where the
information has to be split into several packets which can be lost in
transmission resulting in poor signal quality (\cite{jayant1},
\cite{jayant2}, \cite{yang}, \cite{batllo}).

\subsection{Problem Statement}

We are given a communication system with $k$ channels. Channel $i$
transmits information reliably at rate $\log n_i$ bits per second.
Each channel either succeeds or fails to transmit the information. If
a channel fails, all the information transmitted over the channel is
lost. If a channel succeeds, the received information is assumed to be
correct.

We assume the source to generate integers with uniform distribution,
the result of a quantization process. It is unimportant what the
numbers actually are, and we refer to them by the indices $0$ through
$m-1$ for some $m$. Thus we consider the information to be transmitted
to be an integer $x$ with at most $\lg m$ bits, where $1 \le m \le
n_1n_2\cdots n_k$. An $(n_1,n_2,...,n_k)$--level MSDQ maps $x$ to a
unique $k$-tuple $(i_1,...,i_k)$; component $i_j$ is sent over the
$j$th channel. If all of the channels succeed, then we want to be able
to decode $x$ from the $k$-tuple exactly.  If some of the channels
fail, we want the encoding to minimize the distortion between the
original information and the reconstructed transmission. The system
can be viewed as one encoder $$f:\{0,...,m-1\} \to
\{0,...,n_1-1\}\times\cdots\times\{0,...,n_k-1\}$$ and $2^k-1$
decoders, $(g_0,...,g_{2^k-2})$, each dealing with a unique subset of
successful channels, with at least one succeeding channel. Let $g_0$
be the decoder with all channels succeeding.  The problem we are
interested is describing the rate-distortion tuples $$(\log n_1,\log
n_2,...,\log n_k;0,D_1,...,D_{2^k-2}),$$ where $D_t$ is the distortion
rate of the set of channels represented binary by $t$, a $1$
corresponding to failure. This is a generalization of the notation
used for two channels, where the rate distortion tuples are specified
by $(R_1,R_2; D_0,D_1,D_2)$. Here $R_1$ and $R_2$ are the transmission
rates of the two channels, $D_0$ is the distortion in case both
channels succeed, $D_1$ is the distortion in case of the first channel
failure, and $D_2$ is the distortion in case of the second channel
failure.

Consider the available information in case of channel failure. If some
of the channels fail, the remaining successful channels imply upper
and lower bounds on $x$, namely the largest and the smallest values
among those consistent with the successful components of $x$.  For
example, when $k=2$, $x$ is mapped to a pair $(i_1,i_2)$. If the first
channel fails, we know $x$ is between the smallest and the largest
numbers having second component $i_2$, while if the second channel
fails, $x$ is between the smallest and the largest numbers having
first component $i_1$.  Thus, designing a code to minimize distortion
in case of any $l$ channels failing in a system with source dictionary
of size $m$ and $k$ channels, channel $i$ transmitting reliably at
rate $\log n_i$, is equivalent to the combinatorial problem of putting
numbers $X=\{0,...,m-1\}$ into a $k$-dimensional matrix, dimension $i$
being of size $n_i$, to minimize the difference between the smallest
and the largest number in each full $l$-dimensional submatrix. This
correspondence is shown in Figures \ref{channel_matrix1} and
\ref{channel_matrix2}. In this paper we will be working with the
combinatorial version of the problem.

\begin{figure}[t]
\unitlength1cm
\begin{minipage}[t]{4.5cm}
\begin{picture}(4.5,7.5)
\rput[r](1,6.5){$x$}
\psline{->}(1,6.5)(1.5,6.5)\psframe*(1.5,6)(2.5,7)
\psline{->}(2.5,6.5)(3,6.5)\psline{-}(3,6.8)(3,6.2)
\psline{-}(3,6.8)(3.5,6.8)\rput(3.65,6.8){\small$i_1$}
\psline{-}(3,6.2)(3.5,6.2)\rput(3.65,6.2){\small$i_2$}
\psline[doubleline=true]{->}(2,5.9)(2,4.5)
\psframe(.5,0)(4.5,4)
\rput[r](.4,2){\small$i_1$}\rput[b](3.3,4.05){\small$i_2$}
\psline[linestyle=dotted]{-}(.5,2)(4.5,2)\psline[linestyle=dotted]{-}(3.3,0)(3.3,4)
\rput*[framesep=0pt](3.3,2){$x$}
\end{picture}\par
\caption{Correspondence between an encoding scheme and arrangement of numbers
in a matrix.}
\label{channel_matrix1}
\end{minipage}
\hfill
%3a
\begin{minipage}[t]{10.5cm}
\begin{picture}(10.5,8)
\rput(2.5,5.8){$x$}
\psline{-}(2.7,5.8)(3.2,5.8)\psline{-}(3.2,5.5)(3.2,6.1)
\psline{-}(3.2,6.1)(3.7,6.1)\rput[l](3.75,6.1){\small$x$}
\psline{-}(3.2,5.5)(3.7,5.5)\rput[l](3.75,5.5){\small$x$}
\rput[b](3.25,1){\psframebox{\scriptsize
$\left.\matrix{
0&&&&&&\cr
&1&&&&&\cr
&&2&&&&\cr
&&&&&&\cr
&&&&\ddots&&\cr
&&&&&&\cr
&&&&&&\cr
&&&&&&\cr
&&&&&&15\cr
}\right.$}}
\psline(5.5,6.5)(5.5,.5)
%%3b
\rput(7,5.8){$x=0110$}\psline[linewidth=0.5pt]{-}(7.35,5.5)(7.35,6.1)
\psline{-}(7.9,5.8)(8.4,5.8)\psline{-}(8.4,6.1)(8.4,5.5)
\psline{-}(8.4,5.5)(8.9,5.5)\rput[l](8.95,5.5){\small$01$}
\psline{-}(8.4,6.1)(8.9,6.1)\rput[l](8.95,6.1){\small$10$}
\rput[b](7.75,1.8){\psframebox{\scriptsize
$\left.\matrix{
0&1&2&3\cr\cr
4&5&{\bf6}&7\cr\cr
8&9&10&11\cr\cr
12&13&14&15\cr
}\right.$}}\pscircle(8.05,3.2){.2}
\rput[r](6.4,3.2){\scriptsize$1$}
\rput[b](8.05,4.15){\scriptsize$2$}
\end{picture}
\par
\caption{Examples of encoding schemes and the corresponding matrices:
{\it left} - identical copies of the input number are sent over the channels;
{\it right} - the input number is split into blocks of bits, which are sent
over the channels.}
\label{channel_matrix2}
\end{minipage}
\end{figure}

The problem is also equivalent to minimizing graph bandwidth of a
$k$-fold cartesian product of cliques (Hamming graph) and induced
subgraphs of it. There is a large body of research dedicated to the
bandwidth of various graphs.  There are two possible simplifications
of the Hamming graph bandwidth problem: either small cliques, or few
cliques.  In 1966 Harper \cite{harper2} solved the bandwidth problem
for a $k$-dimensional hypercube, the cartesian product of $k$ cliques
of size 2. Hendrich and Stiebitz \cite{HS} solved the problem for
cartesian product of two cliques of equal size. We propose a vertex
labeling of the cartesian product of an arbitrary number of cliques
of arbitrary (equal) size. To the best of our knowledge, result gives
the best upper bound on the bandwidth of products of more than two
cliques of size greater than 2.

For a survey on the topic of graph bandwidth up to 1982, see
\cite{chinn}. For a recent survey on Harper-type techniques on graphs
see \cite{bezrukov}.  For more information on the subject of graph
bandwidth see West, \cite{west}.

\subsection{Notation and Terminology}
\begin{itemize}
\item{\em Arrangement} -- the inverse of encoding, that is a function from
the cells of the matrix to the numbers to be put in those cells:
$$A:I \to \{1,...,m\},$$ where $I$ is a subset of the product of the sets of
indices $I_1\times\cdots\times I_k$.
\item{\em Slice} -- a full submatrix. An $l$-dimensional slice is a subset of all cells with $k-l$ coordinates fixed:
$$(*,...,*,i_{j_ 1},*,...,*,i_{j_2},*,...,*,i_{j_{k-l}},*,...,*).$$
\item{\em Line} -- a one-dimensional slice:
$$(i_1,i_2,...,*,...,i_k).$$
\item{\em Spread} -- the difference between the largest and the smallest
number in a slice.
\item{\em Maximum spread} of an arrangement, $spread(A)$, -- the maximum over all the spreads in slices of the same fixed dimension.
\item{\em Smalls} -- a plural form of ``the smallest number'', a set of the smallest numbers in a set of slices.
\item{\em Bigs} -- similar to smalls, a plural form of ``the largest number''.
\end{itemize}

\section{Results}

%Given $k$ channels of capacities $\log n_1,\log n_2,...,\log n_k$, and 
%numbers from a set $X=\{x_0,x_1,...,x_{m-1}\}$ to be transmitted over those
%channels, we want to determine the best mapping of $x \in X$ to a $k$-tuple
%$(i_1,i_2,...,i_k)$ so that if some channels fail, we can give as small as
%possible range for $x$ on the receiving end of the transmission. We are
%interested in minimizing the maximum error. This 
%translates into the problem of finding an arrangement of the elements of $X$
%in a $k$-dimensional $n_1 \times n_2 \times ...\times n_k$ matrix that
%minimizes the spread in each projection. The coordinates of that matrix
%in each dimension $i$ can have values from $0$ to $n_i-1$, where $n_i$ is
%the size of dimension $i$.

%Given an encoding, which is an arrangement of $m$ numbers in a 
%$n_1\times\cdots\times n_k$ $k$-dimensional matrix, the error in case of
%failed channels $f_1, f_2,...,f_l$ is the difference between the
%minimum and the maximum numbers, the spread, in the $l$-dimensional projection
%of the successful channels onto the the space of the failed channels:
%$(i_1,...,i_{f_1-1},*,i_{f_1+1},...,i_{f_l-1},*,i_{f_l+1},...,i_k),$
%where $i_j$ is the number transmitted and received over a successful channel.

%In general, an arbitrary set of numbers $\{x_0,x_1,...,x_{m-1}\}$ can be sent 
%over the channels. However, because we can represent those numbers by their
%indices  $\{0,...,m-1\}$, and because
%the proof is based only on the order of the numbers, and not on their
%actual values, we can assume that $X = \{0,1,...,m-1\}$.

We design a technique that provides a lower bound on the maximum
spread in a line in any arrangement of the numbers $X=\{0,...,m-1\}$
in a $k$-dimensional matrix. The technique is constructive, which
allows us to design an algorithm that gives an upper bound.

First, we consider the case of equal channel capacities, so that the
corresponding $k$-dimensional matrix is a cube. We discuss the unequal
channel capacities in Section \ref{unequal}. In addition, the lower
bound and the algorithm are derived for the case of only one channel
failing. In Section \ref{many_channels} we show the results for an
arbitrary number of channels failing. The $k$-channel problem thus
reduces to finding an arrangement of the $m$ numbers in an
$n\times\cdots\times n$ $k$-dimensional matrix that minimizes the
maximum spread in a line.

\begin{figure}
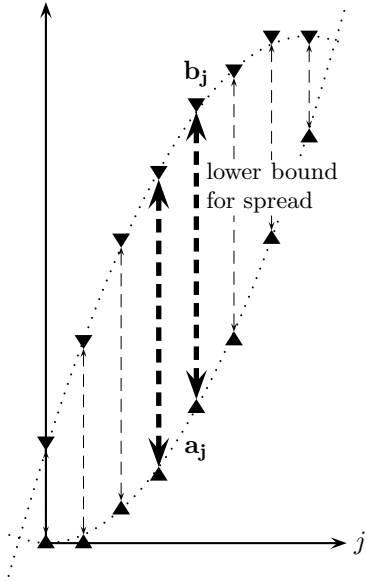

\begin{center}
\psset{xunit=0.5cm, yunit=.45cm}
\pspicture(-1,-1)(10,16)
\psline{<->}(0,16)(0,0)(8,0)\rput[l](8.2,0){$j$}
\pscurve[linestyle=dotted](-1,0.25)(0,0)(1,0.25)(2,1)(3,2.25)(4,4)(5,6.25)(6,9)(7,12.25)(8,16)
\psdots[dotstyle=triangle*, dotsize=5pt 0](0,0)(1,0)(2,1)(3,2)(4,4)(5,6)(6,9)(7,12)
\pscurve[linestyle=dotted](-1,-1)(0,2.75)(1,6)(2,8.75)(3,11)(4,12.75)(5,14)(6,14.75)(7,15)(8,14.75)
\psdots[dotstyle=triangle*, dotangle=180,dotsize=5pt 0](0,3)(1,6)(2,9)(3,11)(4,13)(5,14)(6,15)(7,15)
{\psset{linewidth=.1pt,linestyle=dashed}
\psline{<->}(0,0.25)(0,2.75)
\psline{<->}(1,0.25)(1,5.75)
\psline{<->}(2,1.25)(2,8.75)
\psline[linewidth=2pt]{<->}(3,2.25)(3,10.75)
\psline[linewidth=2pt]{<->}(4,4.25)(4,12.75)
\psline{<->}(5,6.25)(5,13.75)
\psline{<->}(6,9.25)(6,14.75)
\psline{<->}(7,12.25)(7,14.75)
}
\rput[b](4,13.5){$\bf b_j$}
\rput[t](4,3){$\bf a_j$}
\rput[l](4.2,10.5){\psframebox*[framesep=1pt]{\parbox[l]{1.75cm}{\small lower bound for spread}}}
\endpspicture
\end{center}
\caption{The process of derivation of the lower bound on spread.}
\label{seq}
\end{figure}

The idea of the lower bound proof is as follows:
\begin{enumerate}

\item For any possible arrangement $A$ of $X=\{0,...m-1\}$ in the matrix,
consider the sorted (in ascending order) list of smalls in all lines,
$small(A)$. If a number is the smallest one in more than one line,
than it appears in this list more than once. For example, $0$ always
appears $k$ times and any $small(A)$ list starts with $k$ zeros. The
goal is to find a bounding sequence of smalls that is at least as
large elementwise as any such smalls list.  Then the $j$th smallest
number in a line in any arrangement, $small(A)_j$, is at most the
$j$th member of the bounding sequence. Let $\langle a\rangle = \langle
a_1,a_2,...\rangle$ be the bounding sequence; then the following must
hold for all $j$: $$a_j = \max_A\{small(A)_j\}.$$ We will show in
Lemma \ref{lemma1} that there exists an arrangement whose smalls list
realizes the bounding sequence.

\item Similar to the smalls, find a bounding sequence of bigs that is
elementwise at most any bigs list, $big(A)$, produced by any
arrangement. Let $\langle b\rangle = \langle b_1,b_2,...\rangle$ be
the bounding bigs sequence; then the following must hold for all $j$:
$$b_j = \min_A\{big(A)_j\}.$$ Lemma \ref{lemma1} shows that there
exists an arrangement whose bigs list realizes the bounding sequence.

\item Maximum pairwise difference of the bigs and smalls lists of an
arrangement is a lower bound on the maximum spread of the arrangement,
$spread(A)$.  That is, for all $A$ $$\max_j \{big(A)_j-small(A)_j\}\le
spread(A).$$ This statement is known as the Ski Instructor problem and
the proof can be found in \cite{lawler}.

%\begin{proof}
%Let $j = \arg(\max_j (max(A)_j-min(A)_j))$. For any permutation of $min(A)$
%one of the three following things can happen:
%\begin{itemize}
%\item $max(A)_j$ is still paired up with $min(A)_j$, and thus the maximum
%pairwise difference has not decreased;
%\item $max(A)_j$ is paired up with $min(A)_i$ where $min(A)_i\le min(A)_j$,
%but then $max(A)_j-min(A)_j \le max(A)_j - min(A)_i$;
%\item $max(A)_j$ is paired up with $min(A)_k$ where $min(A)_k\ge min(A)_j$,
%but then some element $max(A)_p, p > j$ is paired up with some
%$min(A)_q, q \le j$ by Pigeonhole Principle, and $max(A)_j-min(A)_j\le
%max(A)_p-min(A)_q$.
%\end{itemize}
%\end{proof}

Since for the bounding sequences $\langle b\rangle$ and $\langle
a\rangle$ we have $b_j \le big(A)_j$ and $a_j \ge small(A)_j$ for all
$A$ and $j$, then pairing smaller $b_j$ with smaller $a_j$ gives a lower
bound on the spread for all possible arrangements. For all $A$:
$$\max_j (b_j-a_j) \le \max_j\{big(A)_j-small(A)_j\}\le spread(A).$$ The
process is shown schematically in Figure \ref{seq}.
\end{enumerate}

Thus, the main focus of our proof is finding good bounding
sequences. Consider the smalls sequence.  Suppose we have an initial
segment of the bounding smalls sequence, $\langle
a_1,a_2,...,a_j\rangle$, and are now concerned with the next element
in that sequence.  As we place the elements of $X$ in increasing order
into the cells, the key observation to maximizing the smalls sequence
is that if $x$ is a value in some cell (which can be thought of as the
intersection of $k$ lines), then $x$ is the smallest number in every
line that does not already have an element smaller than $x$. For
example, if we put $x$ into a cell that is an intersection of lines
that do not currently have any elements in them, then $x$ is the
smallest number in $k$ lines, and thus appears in the smalls sequence
$k$ times.  On the other hand, if all $k$ lines already have numbers
less than $x$, then it does not appear in the smalls sequence at all,
and the next candidate for the sequence member is at least $x+1$.  In
general, if $s$ out of $k$ lines have elements less than $x$, then $x$
appears in the smalls sequence exactly $k-s$ times.  Therefore, to
maximize the next element of the smalls sequence we want to put $x$
into a cell that is an intersection of as many as possible lines that
already have elements lass than $x$. Given a choice, we would also
like to put $x$ in a cell that reduces the number of lines without
smaller values for the subsequent elements. An example of such
placement is shown in Figure
\ref{place}. We now demonstrate the lower bound proof on some special
cases.

\begin{figure}
\begin{center}
\begin{tabular}{|p{.5cm}|p{.5cm}|p{.5cm}|}
\hline
&&\cr
$x_1$ 	&  $x_2$  & * \cr
\hline
&&\cr
*	& *	& \cr
\hline
&&\cr
&& \cr
\hline
\end{tabular}
\end{center}
\caption{Let $x_1$ and $x_2$ be the elements less than $x$ that already have
been placed. All cells marked $*$ are intersections of two lines, one
of which already has a number smaller than $x$. However putting $x$ in
the cell below $x_1$ or $x_2$ will produce one cell that is an
intersection of two lines, both of which have a smaller number. Thus
we favor those over the cell to the right of $x_2$, since placing $x$
there results only in cells with at most one smaller number in their
intersection.}
\label{place}
\end{figure}

\subsection{The Completely Filled Cube}
\label{cube}

Assume that the matrix is cube, so $n_i=n$ for all $i$ and the number
of elements to be placed in the matrix is $m=n^k$. This corresponds to
all channels capacities being equal and the numbers to be transmitted
over the channels have number of bits up to the sum of the number of
bits that can be transmitted over each channel. This is the most
commonly used setup in practice, especially in the context of packet
switched network. From the rate-distortion point of view, this
corresponds to tuples of type $(\log n, \log n,...,\log n; 0,
D_1,...,D_{2^k-1})$.

First we show that for a completely filled matrix, it is sufficient to
restrict our attention to a special kind of arrangements.

\begin{definition}
Extending the definition in \cite{FTW}, an arrangement $A$ is {\em
monotonic} if for for any line $(i_1,...,*,...,i_k)$ if $p > q$ then
$$A(i_1,...,p,...,i_k) > A(i_1,...,q,...,i_k).$$
\end{definition}

\begin{lemma}
\label{monotonic_lemma}
The maximum spread in a completely filled cube is minimized by a
monotonic arrangement.
\end{lemma}
\begin{proof}
\begin{figure}
\begin{center}
{\psset{unit=.5}
\begin{pspicture}(0,0)(12,12)
\psframe(0,0)(12,12)
\psframe(0,3)(12,4)\psframe(0,8)(12,9)
{\psset{fillstyle=solid,fillcolor=lightgray}
\psframe(0,8)(5,9)\psframe(0,3)(2,4)\psframe(3,3)(5,4)
}
\psframe(2,0)(3,12)\psframe(5,0)(6,12)\psframe(9,0)(10,12)
\rput(2.5,12.5){$s$}\rput(5.5,12.5){$j$}\rput(9.5,12.5){$t$}
\rput(2.5,8.5){$c_s$}\rput(5.5,8.5){$c_s$}\rput(9.5,8.5){$c_t$}
\pscurve{->}(2.5,9)(4,9.5)(5.5,9)
\pscurve{->}(9.5,9)(9.25,9.75)(8.5,9.5)\pscurve{->}(9.5,9)(9.75,9.75)(10.5,9.5)
\rput(2.5,3.5){$d_s$}\rput(5.5,3.5){$d_t$}\rput(9.5,3.5){$d_t$}
\pscurve{<-}(5.5,4)(7.5,4.5)(9.5,4)
\pscurve{->}(2.5,4)(2.25,4.75)(1.5,4.5)\pscurve{->}(2.5,4)(2.75,4.75)(3.5,4.5)
\end{pspicture}
}
\end{center}
\caption{Rearrangement in one coordinate causes the spread in another
coordinate to become $d_t - c_s$. Shaded are the $c$'s less than $c_s$
but not equal to $c_t$ and $d$'s that are less than $d_t$ and not
equal to $d_s$.}
\label{monotonic}
\end{figure}
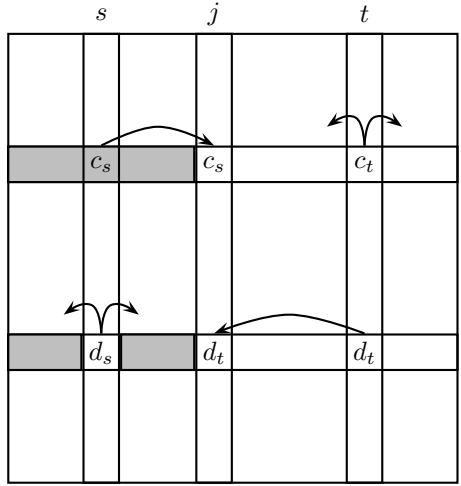
We first show that given any arrangement completely filling the
matrix, rearranging the numbers to become monotonic in one coordinate
does not increase the overall spread. It is obvious that rearranging
the numbers in any way within the same line does not change the spread
in that line, thus the rearrangement for full monotonicity within a
coordinate does not change the spread in that coordinate.  Suppose the
spread has increased for some coordinate. The situation is described
in Figure \ref{monotonic}.  Suppose after the rearrangement the
maximum spread in that coordinate is in line $j$ and is $d_t - c_s$
(where $d_t$ was in line $t$ before the rearrangement, and $c_s$ was
in line $s$). Then $$d_t - c_t < d_t - c_s, \mbox{ thus }d_s < d_t,
\mbox{ and}$$ $$d_s - c_s < d_t - c_s, \mbox{ thus }c_s < c_t.$$ Then
there are $j-2$ (since $d_t$ and $c_s$ are now in line $j$) $d$'s less
than $d_t$ and not equal to $d_s$. There are $j-1$ $c$'s less than
$c_s$ and not equal to $c_t$. Therefore, by Pigeonhole Principle,
there exists $c_p < c_s$ that was paired up with $d_p > d_t$ before
the rearrangement. But then $d_p - c_p > d_t - c_s$.

We have proved that rearranging the numbers in one coordinate line
after line does not increase the spread in the matrix. Gale and Karp
\cite{GK} show that if the arrangement was monotonic in one coordinate
then it will remain so after the number are rearranged to become
monotonic in another coordinate. Thus the matrix can be rearranged to
have a fully monotonic arrangement one coordinate at a a time, one line
at a time, without increasing the spread.
\end{proof}

Thus, it is sufficient to consider only monotonic arrangements.

We can now define the arrangement that produces a bounding smalls
sequence for the cubic matrix. In fact, this arrangement produces a
bounding smalls sequence for a more general class of completely filled
rectangular matrices, cube being a special case.

\begin{definition}
\label{def1}
\end{definition}
A {\em herringbone arrangement} of a $k$-dimensional $n_1\times
n_2\times\cdots\times n_k$ completely filled matrix is defined
inductively as follows.  Assign an arbitrary order to the coordinates
of the system $\langle i_1, i_2,...,i_k\rangle$.  A herringbone
arrangement of $0\times\cdots\times 0$ $k$-dimensional matrix is
empty.  A herringbone arrangement of a $0$-dimensional matrix is the
number $0$ placed in a single cell.  Given a herringbone arrangement
of $t_1\times t_2\times\cdots\times t_k$ $k$-dimensional matrix (that
is the cells of the matrix are filled up to the coordinate $t_i-1$ in
dimension $i$), we define a larger herringbone arrangement
inductively:
\begin{itemize}
\item project the existing arrangement onto the $(k-1)$-dimensional 
slices adjacent to the existing arrangement,
\item calculate the $(k-1)$-dimensional volume of each projection,
\item recursively fill the largest volume projection (using coordinate order
to break ties) with the herringbone arrangement for $k-1$ dimensions.
\end{itemize}

\begin{figure}
\begin{center}
\begin{pspicture}(0,0)(3,4)
\rput(-4,1){\begin{pspicture}(0,-3)(3,0)
\psset{dimen=middle}
\psframe(0,-3)(3,0)
\psline(.5,0)(.5,-.5)
\psline(0,-.5)(1,-.5)
\psline(1,0)(1,-1)
\psline(0,-1)(1.5,-1)
\psline(1.5,0)(1.5,-1.5)
\psline(0,-1.5)(2,-1.5)
\psline(2,0)(2,-2)
\psline(0,-2)(2.5,-2)
\psline(2.5,0)(2.5,-2.5)
\psline(0,-2.5)(3,-2.5)
\end{pspicture}}
\rput(0,2.5){\begin{pspicture}(0,0)
\psset{dimen=middle}
\pspolygon(0,-3)(3,-3)(4,-2)(4,-1.5)(3,-2.5)(0,-2.5)(0,-3)
%pink
\psframe(0,-2.5)(3,0)
\psline(0,-1)(1,-1)
\psline(0,-1.5)(1.5,-1.5)
\psline(0,-2)(2,-2)
\psline(0,-2.5)(2.5,-2.5)
%red
\pspolygon(0,0)(1,1)(3.5,1)(2.667,.167)(3.167,.167)(3.167,-2.33)(3,-2.5)(2.5,-2.5)(2.5,-2)(2,-2)(2,-1.5)(1.5,-1.5)(1.5,-1)(1,-1)(1,-.5)(0.5,-.5)(0.5,0)(0,0)
\psframe(0,0)(0.5,-.5)
\psline(0.666,.666)(1.666,.666)
\psline(0.5,.5)(2,.5)
\psline(0.333,.333)(2.333,.333)
\psline(0.167,.167)(2.667,.167)
\psline(0.5,-.5)(0.5,0)(0.667,.167)
\psline(1,-.5)(1,0)(1.167,.167)
\psline(1.5,-1)(1.5,0)(1.667,.167)
\psline(2,-1.5)(2,0)(2.167,.167)
\psline(2.5,-2)(2.5,0)(2.667,.167)
%lightblue
\pspolygon(3.167,.167)(4,1)(4,-1.5)(3.167,-2.33)(3.167,.167)
\psline(4,0)(3.666,-.334)
\psline(4,-.5)(3.5,-1)
\psline(4,-1)(3.333,-1.667)
%blue
\pspolygon(2.667,.167)(2.833,.333)(2.333,.333)(2.5,.5)(2,.5)(2.166,.666)(1.666,.666)(1.833,.833)(0.833,.833)(1,1)(4,1)(4,.5)(3.666,.167)(3.666,-.334)(3.5,-.5)(3.5,-1)(3.333,-1.167)(3.333,-1.667)(3.167,-1.834)(3.167,-2.167)(3.167,.167)(2.667,.167)
\pspolygon(0.833,.833)(1,1)(1.5,1)(1.333,.833)(0.833,.833)
\pspolygon(3.833,.833)(4,1)(4,.5)(3.833,.333)(3.833,.833)
\psline(1.5,1)(1.333,.833)
\psline(2,1)(1.833,.833)
\psline(2.5,1)(2.166,.666)
\psline(3,1)(2.5,.5)
\psline(3.5,1)(2.833,.333)
\psline(2.833,.333)(3.333,.333)(3.333,-1.167)
\psline(3,.5)(3.5,.5)(3.5,-.5)
\psline(3.166,.666)(3.666,.666)(3.666,.167)
\psline(3.333,.833)(3.833,.833)(3.833,.333)
\psline(0,0)(3,0)(3,-3)
\psline(3,0)(4,1)
\end{pspicture}}
\rput(7,2){\epsfxsize 2 in \epsfbox{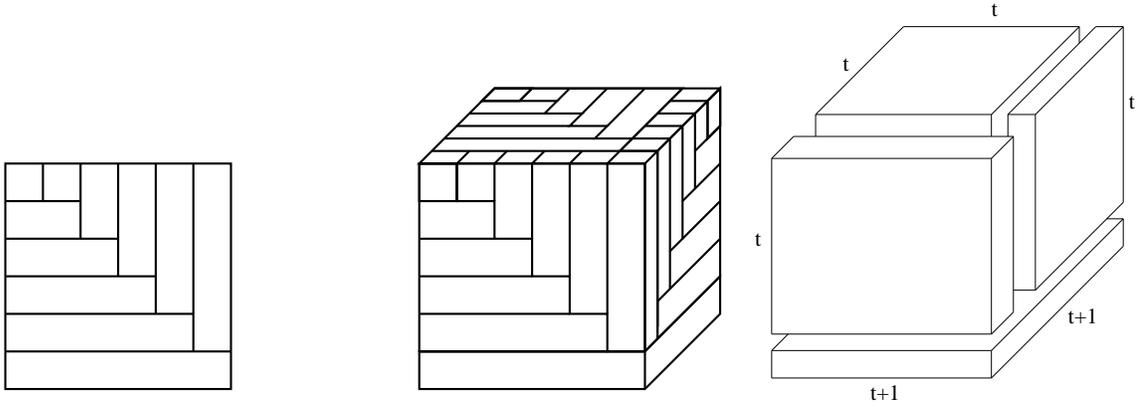}}
\end{pspicture}
\end{center}
\caption{An example of herringbone arrangements in 2 and 3-dimensional matrices.}
\label{cube_HB}
\end{figure}

Examples of $2$ and $3$-dimensional arrangements are shown in Figure
\ref{cube_HB}. The name ``herringbone arrangement'' is due to the
herringbone-like pattern seen clearly in two dimensions.  We denote
the element in the cell $(i_1,...,i_k)$ of the $k$-dimensional
herringbone arrangement by $HB_k(i_1,...,i_k)$. Let $i_{\max} =
i_p$. If there is more than one coordinate with the maximum value,
take the largest coordinate. Then $$HB_k(i_1,...,i_k)=
(i_{p}+1)^{(p-1)}\cdot i_{p}^{(k-p+1)} +
HB_{k-1}(i_1,...,i_{p-1},i_{p+1},...,i_k).$$ The last equality follows
from the recursive definition of the herringbone arrangement. The
herringbone arrangement fills the matrix in layers, the maximum
coordinate value indicates which layer of the arrangement a cell is
in. Thus the value in a cell is in the $i_p$'th layer, the first
$i_p-1$ layers being completely filled and the element is within a
$(k-1)$-dimensional submatrix, recursively filled with the herringbone
arrangement.

\begin{lemma}
\label{lemma1}
The herringbone arrangement of values in a $k$-dimensional matrix maximizes
the smalls sequence -- the ascending list of lines' smallest numbers  -- for that matrix.
\end{lemma}
\begin{proof}
The proof is a generalization of Harper's proof of the main theorem in
\cite{harper1}. We use induction on $k$ and the largest dimension
size, $n_{\max}=\max_i \{n_i\}$.

%If $k=0$, $n_{\max}>0$, then we have only one cell and one
%element. Any arrangement is a herringbone arrangement and any
%arrangement produces the same minima sequence of one element.\\ If $k
%> 0$ and $n_{\max}=1$, then there are $k$ lines of size one, and the
%$a_j$ sequence consists of $k$ zeroes, which is achieved by any way of
%putting $0$ in the only available cell.

The base cases of $k=0$ and $n_{max}=1$ are trivial.

Suppose we have a matrix with the largest dimension size $n_{\max}$
(largest coordinate value $n_{\max}-1$).  By the induction hypothesis,
the herringbone arrangement maximizes the smalls sequence in the
$k$-dimensional matrix up to the coordinate value $n_{\max}-2$ in
every dimension, that is, it maximizes the initial segment of the
smalls sequence for the entire matrix.

Consider the smallest element $x$ which has not yet been used in the
arrangement. As we have noted before, every cell in the matrix is an
intersection of $k$ lines. We shall call a line {\em protected} if it
has a smallest number in it. Since we are placing numbers in the
increasing order, this means a line is protected if it has any number
in it. When we put $x$ in any cell, it will be the smallest number in
any unprotected line in its intersection. Thus the goal is to put $x$
into a cell that is an intersection of as many as possible protected
lines. However, since we have a complete herringbone arrangement of a
smaller cube matrix, any free cell has at most one protected line in
its intersection. The cells that have one protected line are precisely
the cells that lie in the lines that intersect a face of the existing
herringbone arrangement. Consider now all the lines that intersect one
face. After placing the first element in any of these lines, there
always exists a cell that is an intersection of at least two protected
lines.  Thus, once started, one must stay with the same face to ensure
larger elements in the smalls list. Notice, that the cells that are
being filled are exactly a $(k-1)$-dimensional projection of a face of
a herringbone arrangement, that is a $(k-1)$-dimensional matrix. Thus
by induction hypothesis it is filled with a herringbone arrangement.

The question that remains is which one of the faces should one start
with.  Notice, that one of the properties of the herringbone
arrangement is that at any point the size of the available faces
differs by at most $1$ in any dimension, and they can differ in at
most one dimension. Suppose we have one face $F_1$ of size
$t_1\times\cdots\times t_i\times\cdots\times t_k$ and another face
$F_2$ of size $t_1\times\cdots\times t_i+1\times\cdots\times t_k$.  It
is easy to see that the smalls sequence of the $F_1$ face agrees with
the initial segment of the smalls sequence of the $F_2$ face, given
they are filled with the same numbers. The next element of the smalls
sequence of the $F_2$ face appears there exactly $k-2$ times. However,
after filling the $F_1$ face we must start a new face, and the next
(same) element in the sequence would appear there $k-1$ times, thus in
this case we get a smaller element in the smalls sequence.  Therefore,
to maximize the smalls sequence we must first fill the larger volume
face projections.

The above arguments produce, by Definition \ref{def1}, a herringbone
arrangement.
\end{proof}

We are now ready to give a lower bound on the spread in a completely filled 
cube.
\begin{theorem}
\label{thrm1}
The spread in a completely filled cube is at least
$$n^k - 1 - 
\left\lfloor{\left({kn^{k-1}+2\over 2k}\right)^{k\over k-1}}\right\rfloor - 
\left\lfloor{\left({kn^{k-1}\over2k}\right)^{k\over k-1}}\right\rfloor.$$
%$$\left.\matrix{
%n^k - 1 - n\left({\left({n+1\over 2}\right)^{k-1} - 1}\right),\hfill&
%\mbox{ if $n$ is odd,}\hfill\cr\cr
%n^k+n-2-\left({n\over2}\right)^{k-1\over2}\left({(n+1)\left({n+2\over2}\right)^{k-1\over2}-2}\right),\hfill&
%\mbox{ if $n$ is even, and $k$ is odd,}\hfill\cr\cr
%n^k+n-2-\left({n\over2}\right)^{k-2\over2}\left({n+2\over2}\right)\left({n\left({n+2\over 2}\right)^{k-2\over2}-1}\right),\hfill&
%\mbox{ if both $n$ and $k$ are even.}\hfill\cr
%}\right.
%$$
\end{theorem}
\begin{proof}
By Lemma \ref{lemma1}, herringbone arrangement of a completely filled
cube maximizes the smalls sequence. A process similar to the
derivation of the bounding smalls sequence creates a bounding bigs
sequence.  For the bigs sequence, however, we start instead with the
largest element in the cell with the largest coordinate value and work
our way downward.

Since for any arrangement of the elements in the matrix, we know the
bounding sequences $a_j$ and $b_j$, the spread for any arrangement is
at least $\max_j\{b_j-a_j\}$.  The closed formula for $a_j$ is
unusably complicated. However, consider the case of $j=kt^{k-1}$ for
some $t$. In this case $a_j$ is the minimum in the first line after
filling a subcube with sides of size $t$, that is $a_j = t^k =
(j/k)^{k\over k-1}$. Thus $\lfloor{(j/k)^{k\over k-1}}\rfloor$ is a
crude overestimate of any $a_j$ (rounding up to the closest $t^k$)
that coincides with $a_j$ in infinitely many values. The sequence
$b_j$ is complementary of $a_j$.  There are $kn^{k-1}$ lines in a
$k$-dimensional cube, therefore there are $kn^{k-1}$ elements in the
smalls and bigs sequences, therefore the index complimentary to $j$ in
the sequence is $kn^{k-1}-j+1$ and $$b_j = n^k-1-a_{kn^{k-1}-j+1} \ge
n^k - 1-\lfloor((kn^{k-1}-j+1)/k)^{k\over k-1}\rfloor.$$ Since the
sequences $b_j$ and $a_j$ are complimentary and $a_j$ is convex, then
$b_j$ is concave and $\max_j\{b_j-a_j\}$ is achieved in the middle of
the sequence, that is, when $j = kn^{k-1}/2$.
\begin{eqnarray*}
\max_j\{b_j-a_j\} &\ge& 
b_{kn^{k-1}/2} - a_{kn^{k-1}/2}\cr &=&
\left(n^k - 1 - 
\left\lfloor{\left({kn^{k-1}-{kn^{k-1}\over2}+1\over k}\right)^{k\over k-1}}\right\rfloor\right) -
\left(\left\lfloor{\left({kn^{k-1}\over2k}\right)^{k\over k-1}}\right\rfloor\right)\cr
& = & n^k - 1 -
\left\lfloor{\left({kn^{k-1}+2\over 2k}\right)^{k\over k-1}}\right\rfloor - 
\left\lfloor{\left({kn^{k-1}\over2k}\right)^{k\over k-1}}\right\rfloor\cr
\end{eqnarray*}
\end{proof}

Note, that this lower bound is weak. It is not sufficient to merely
find the maximum difference between the ordered minima and maxima
sequences. There are more constraints that apply to the matching up of
the sequences that can give a higher lower bound. We will discuss some
of them later.

\vskip .1 in
{\bf The algorithm.}
\vskip .05 in

Herringbone arrangements and any symmetric combinations of smalls and
bigs herringbone arrangements are monotonic.  The idea of the
algorithm is to put the two complimentary herringbone arrangements
together, without increasing the spread. Consider all the ways of
merging the two arrangements in a cube. Assuming that the smalls
arrangement starts at the $(0,0,...,0)$ corner, and the bigs
arrangement starts at the $(n-1,...,n-1)$ corner, the possibilities
are defined by the order of the coordinates in building each
arrangement. Thus there are $k!$ possibilities.  First, assume for now
that we can literally merge the two herringbone arrangements by
putting two numbers in every cell of the matrix. In every line the
smallest number will be taken from the smalls herringbone arrangement
and the largest--from the bigs arrangement. Consider any line in the
cube and the corresponding smallest and largest numbers that are
defined by the merging permutation. Similar to an earlier argument,
the maximum difference between the smallest and the largest numbers in
a line occurs in the middle lines, that is the lines of the type
$(\lceil(n-1)/2\rceil,...,*,...,\lfloor(n- 1)/2\rfloor)$, we shall
call it $Pr_p$ if $p$ is the non-fixed coordinate.  Thus to find the
best permutation we calculate the following: {\small
\begin{eqnarray}
\min_{\pi\in Perm}\: {\max_{1\le p\le k}
{\{HB_{\max}(Pr_p) - HB_{\min}(Pr_p)\}}}&=&
\min_{\pi\in Perm}\: {\max_{1\le p\le k}
{\{n^k-1-(HB_{\min}(Pr_{\pi(p)})+HB_{\min}(Pr_p))\}}}
\nonumber\\
&=&n^k-1-\max_{\pi\in Perm}\: {\min_{1\le p\le k}
{\{HB_{\min}(Pr_{\pi(p)})+HB_{\min}(Pr_p)\}}}\nonumber\\
&=&n^k-1-(HB_{\min}(Pr_1)+HB_{\min}(Pr_k))\label{eq1}
\end{eqnarray}
}

To see why equation \ref{eq1} is true let's look at the smallest
number in $Pr_p$ of the herringbone arrangement as a function of
$p$. We shall show the calculations for odd $n$. The algebra for even
$n$ is similar, and the result is the same.
\begin{eqnarray*}
HB_{\min}(Pr_p) &=&
HB\Bigl({n-1\over2},...,0,...,{n-1\over2}\Bigr), \mbox{ where 0 is in coordinate }p
\\
&=& \left({n+1\over2}\right)^{k-1}\left({n-1\over2}\right)
+ \left({n+1\over2}\right)^{k-2}\left({n-1\over2}\right)
+... + \left({n+1\over2}\right)^p\left({n-1\over2}\right)
\\
&&+\left({n+1\over2}\right)^{p-2}\left({n-1\over2}\right)^2
+\left({n+1\over2}\right)^{p-3}\left({n-1\over2}\right)^2
+...+\left({n-1\over2}\right)^2
\\
&=&\left({n-1\over2}\right)
\sum_{i=p}^{k-1}{\left({n+1\over2}\right)^i} +
\left({n-1\over2}\right)^2
\sum_{i=0}^{p-2}{\left({n+1\over2}\right)^i}
\\
&=&\left({n-1\over2}\right)\left({n+1\over2}\right)^p
{\left({n+1\over2}\right)^{k-p}-1\over
\left({n-1\over2}\right)} + 
\left({n-1\over2}\right)^2
{\left({n+1\over2}\right)^{p-1}-1\over
\left({n-1\over2}\right)}
\\
&=&\left({n+1\over2}\right)^p\Bigl(
\left({n+1\over2}\right)^{k-p}-1\Bigr)+
\left({n-1\over2}\right)\Bigl(
\left({n+1\over2}\right)^{p-1}-1\Bigr)
\\
&=&\left({n+1\over2}\right)^k - 
\left({n-1\over2}\right) - 
\left({n+1\over2}\right)^{p-1}
\end{eqnarray*}

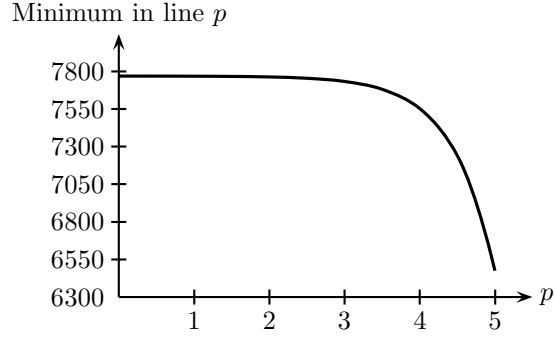
\begin{figure}
\begin{center}
\begin{pspicture}(0,-.2)(5,3.5)
{\psset{yunit=.002}
\savedata{\mydata}[
{{0,1470.83},{.5,1470.6},{1,1470},{1.5,1468.55},{2,1465},{2.5,1456.3},{3,1435},
{3.5,1383},{4,1255},{4.5,942},{5,175}}]
\dataplot[plotstyle=curve,linewidth=1.2pt]{\mydata}
}
\psline{->}(0,0)(5.5,0)\rput[l](5.6,0){$p$}
\multiput(1,0)(1,0){5}{\psline(0,-.1)(0,.1)}
\rput[t](1,-.2){$1$}\rput[t](2,-.2){$2$}\rput[t](3,-.2){$3$}\rput[t](4,-.2){$4$}\rput[t](5,-.2){$5$}
\psline{->}(0,0)(0,3.5)\rput[b](0,3.6){Minimum in line $p$}
\multiput(0,.5)(0,.5){6}{\psline(-.1,0)(.1,0)}
\rput[r](-.2,0){$6300$}\rput[r](-.2,.5){$6550$}\rput[r](-.2,1){$6800$}\rput[r](-.2,1.5){$7050$}\rput[r](-.2,2){$7300$}\rput[r](-.2,2.5){$7550$}\rput[r](-.2,3){$7800$}
\end{pspicture}
\end{center}
\caption{The minimum in line $p$ of the Herringbone arrangement as a function
of $p$, shown here for $n=11$ and $k=5$.}
\label{HB}
\end{figure}
This function is exponential in $p$, with a negative coefficient, thus it
is minimal at $p=k$ (see Figure \ref{HB}). The minimum is so small
relative to the rest of the function, that
$HB(Pr_k)+HB(Pr_i) < HB(Pr_p)+HB(Pr_q)$
for any $i$ and $p,q \ne k$, since even
$$(HB(Pr_k)+HB(Pr_1))-(HB(Pr_{k-1})+HB(Pr_{k-1}))=
-\left({n+1\over2}\right)^{k-1}+
2\left({n+1\over2}\right)^{k-2} -1 < 0.$$
Thus for any permutation $\pi$, 
\begin{eqnarray*}
\min_{1\le p\le k}{\{ HB_{\min}(Pr_{\pi(p)})+HB_{\min}(Pr_p)\} }=\min\{ &
HB_{\min}(Pr_k)+HB_{\min}(Pr_{\pi(k)}),\\
& HB_{\min}(Pr_k)+HB_{\min}(Pr_{\pi^{-1}(k)})\},
\end{eqnarray*}
and the maximum over all permutations $\pi$ is
$$HB_{\min}(Pr_1)+HB_{\min}(Pr_k).$$
This means that one of the best permutations is the reverse permutation, and
the spread achieved by merging the minima and the maxima sequences using
the reverse ordering of the coordinates is
$$n^k-1-(HB_{\min}(Pr_1)+HB_{\min}(Pr_k)) = 
n^k - 1 - n\left({\left({n+1\over2}\right)^{k-1} - 1}\right)$$
when $n$ is odd.
When $n$ is even, the middle lines are of the type
$(i_1,...,*,...,i_k)$ where all the coordinates
equal $\lfloor (n-1)/2\rfloor$ or $\lceil (n-1)/2\rceil$. 
Since the arrangement for the maxima sequence uses the reverse order of the
coordinates,
$$HB_{\max}(i_1,...,*,...,i_k)=n^k-1-HB_{\min}(n-i_k-1,...,*,...,n-i_1-1).$$
Thus the maximum difference occurs in lines with the first half of the
coordinates
being $\lceil(n-1)/2\rceil$ (ignoring the $*$), and the last half of the
coordinates being $\lfloor(n-1)/2\rfloor$. (If $k$ is even, then if $*$ is in
the first half of the coordinate values, there are more floors than ceilings,
and if $*$ is in the last half, then there are more ceilings than floors).
Thus the lower bound on the spread is
$$
HB_{\max}(\underbrace{\left\lceil{n-1\over 2}\right\rceil,...,\left\lceil{n-1\over 2}\right\rceil}_{\lceil(k-1)/2\rceil},\underbrace{\left\lfloor{n-1\over 2}\right\rfloor,...,\left\lfloor{n-1\over 2}\right\rfloor}_{\lfloor(k-1)/2\rfloor},n-1) -
HB_{\min}(\underbrace{\left\lceil{n-1\over 2}\right\rceil,...,\left\lceil{n-1\over 2}\right\rceil}_{\lceil(k-1)/2\rceil},\underbrace{\left\lfloor{n-1\over 2}\right\rfloor,...,\left\lfloor{n-1\over 2}\right\rfloor}_{\lfloor(k-1)/2\rfloor},0) =
$$

$$ 
n^k-1-(HB_{\min}(0,\underbrace{\left\lceil{n-1\over 2}\right\rceil,...,\left\lceil{n-1\over 2}\right\rceil}_{\lfloor(k-1)/2\rfloor},\underbrace{\left\lfloor{n-1\over 2}\right\rfloor,...,\left\lfloor{n-1\over 2}\right\rfloor}_{\lceil(k-1)/2\rceil}) + 
HB_{\min}(\underbrace{\left\lceil{n-1\over 2}\right\rceil,...,\left\lceil{n-1\over 2}\right\rceil}_{\lceil(k-1)/2\rceil},\underbrace{\left\lfloor{n-1\over 2}\right\rfloor,...,\left\lfloor{n-1\over 2}\right\rfloor}_{\lfloor(k-1)/2\rfloor},0) = 
$$

If $n$ is even, this equals to

$$
n^k-1-(HB_{\min}(0,\underbrace{{n\over2},...,{n\over2}}_{\lfloor(k-1)/2\rfloor},\underbrace{{n-2\over2},...,{n-2\over2}}_{\lceil(k-1)/2\rceil}) + 
HB_{\min}(\underbrace{{n\over2},...,{n\over2}}_{\lceil(k-1)/2\rceil},\underbrace{{n-2\over2},...,{n-2\over2}}_{\lfloor(k-1)/2\rfloor},0))= 
$$

$$ 
n^k-1-\Biggl(\left({n\over2}\right)^{\left\lceil{k+1\over 2}\right\rceil}\sum_{i=1}^{\left\lfloor{k-1\over 2}\right\rfloor}{\left({\left({n\over2}\right)+1}\right)^{i}} +
\left({n-2\over2}\right)\sum_{i=1}^{\left\lceil{k-1\over 2}\right\rceil}{\left({\left({n-2\over2}\right)+1}\right)^{i}} + 
$$ $$
\left({n\over2}\right)^{\left\lfloor{k+1\over 2}\right\rfloor+1}\sum_{i=0}^{\left\lceil{k-1\over2}\right\rceil-1}{\left({\left({n\over2}\right)+1}\right)^i}+\left({n-2\over2}\right)^2\sum_{i=0}^{\left\lfloor{k-1\over2}\right\rfloor-1}{\left({\left({n-2\over2}\right)+1}\right)^i}\Biggr) = 
$$

$$
n^k-1-\Biggl(\left({n\over2}\right)^{\left\lceil{k+1\over2}\right\rceil}\left({n+2\over2}\right){\left({n+2\over2}\right)^{\left\lfloor{k-1\over2}\right\rfloor}-1\over \left({n+2\over2}\right)-1} +
\left({n-2\over2}\right)\left({n\over2}\right){\left({n\over2}\right)^{\left\lceil{k-1\over2}\right\rceil}-1\over \left({n\over2}\right)-1} +
$$ $$
\left({n\over2}\right)^{\left\lfloor{k+1\over2}\right\rfloor+1}{\left({n+2\over2}\right)^{\left\lceil{k-1\over2}\right\rceil}-1\over \left({n+2\over2}\right)-1} +
\left({n-2\over2}\right)^2{\left({n\over2}\right)^{\left\lfloor{k-1\over2}\right\rfloor}-1\over \left({n\over2}\right)-1}\Biggr) =
$$

$$
n^k-1-\Biggl(\left({n\over2}\right)^{\left\lceil{k-1\over2}\right\rceil}\left({n+2\over2}\right)\left({\left({n+2\over2}\right)^{\left\lfloor{k-1\over2}\right\rfloor}-1}\right)+
\left({n\over2}\right)\left({\left({n\over2}\right)^{\left\lceil{k-1\over2}\right\rceil}-1}\right)+
$$ $$
\left({n\over2}\right)^{\left\lfloor{k+1\over2}\right\rfloor}\left({\left({n+2\over2}\right)^{\left\lceil{k-1\over2}\right\rceil}-1}\right)+
\left({n-2\over2}\right)\left({\left({n\over2}\right)^{\left\lfloor{k-1\over2}\right\rfloor}-1}\right)\Biggr)
$$

When $k$ is odd, this simplifies to
$$
n^k+n-2-\left({n\over2}\right)^{k-1\over2}\left({(n+1)\left({n+2\over2}\right)^{k-1\over2}-2}\right),
$$

and when $k$ is even, this simplifies to
$$
n^k+n-2-\left({n\over2}\right)^{k-2\over2}\left({n+2\over2}\right)\left({n\left({n+2\over2}\right)^{k-2\over2}-1}\right).
$$

%Notice, that in all cases the lower bound on the spread is $n^k-({n^k\over2^{k-1}}+o(n^{k-1}))$. 
%This quick estimate of $n^k-{n^k\over2^{k-1}}$ for the lower
%bound on the spread of a completely filled cube comes from the following
%intuitive argument. Divide the matrix into $2^k$ equal size pieces by halving
%each dimension. Consider the piece with the smallest numbers in it: the largest
%number there is at least ${n^k\over2^k}$. Consider the piece with the largest
%numbers: the smallest value there is at most $n^k-{n^k\over2^k}$. Since those
%two pieces touch corners, the spread is at least $n^k-{n^k\over2^{k-1}}$.

We have shown that the merging of the two Herringbone arrangements, the minima
and the maxima, gives the spread of:
$$\left.\matrix{
n^k - 1 - n\left({\left({n+1\over 2}\right)^{k-1} - 1}\right),\hfill&
\mbox{ if $n$ is odd,}\hfill\cr\cr
n^k+n-2-\left({n\over2}\right)^{k-1\over2}\left({(n+1)\left({n+2\over2}\right)^{k-1\over2}-2}\right),\hfill&
\mbox{ if $n$ is even, and $k$ is odd,}\hfill\cr\cr
n^k+n-2-\left({n\over2}\right)^{k-2\over2}\left({n+2\over2}\right)\left({n\left({n+2\over 2}\right)^{k-2\over2}-1}\right),\hfill&
\mbox{ if both $n$ and $k$ are even.}\hfill\cr
}\right.
$$

However, we cannot exactly merge the two Herringbone arrangements.
We now present an algorithm that combines the two arrangements and preserves
the spread calculated for the merging. Since the lower bound and the merging
bound coincide for the two dimensional matrices, the algorithm is optimal for
that case. In general, however, it is not optimal and just one of the possible
generalizations of the two-dimensional case.

\vskip .1 in
\noindent{\bf Algorithm HERRINGBONE:}
\begin{enumerate}
\item Fill the initial diagonal half of the matrix
$(i_1,...,i_k), \sum_{j=1}^k{i_j}\le \lfloor k{{n-1}
\over 2}\rfloor$ up to and including the bisecting hyperplane
perpendicular to the main diagonal with the herringbone arrangement
for the minima sequence.
\item Fill the rest of the matrix with the herringbone arrangement for the
maxima sequence, skipping the cells already filled.
\end{enumerate}

\begin{figure}
\begin{center}
\epsfysize 2 in \epsfbox{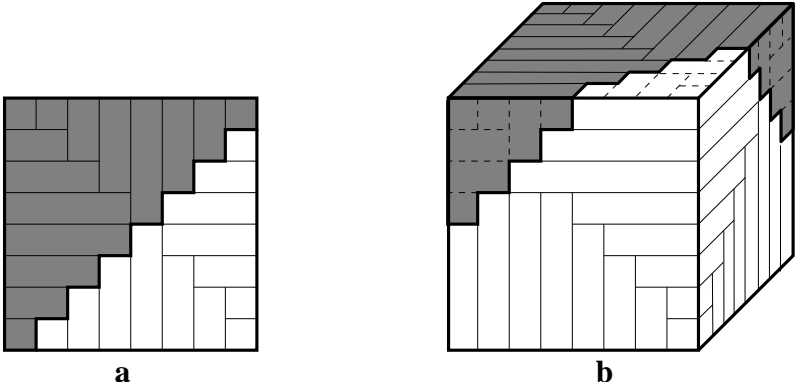}
\caption{The pattern created by the algorithm in (a) two and (b) three 
dimensions. The shaded part
is filled with the herringbone arrangement for minima sequence, the other
half is filled with the arrangement for maxima sequence.}
\label{alg2}
\end{center}
\end{figure}

\begin{theorem}
HERRINGBONE produces an arrangement of a $k$-dimensional cube with
dimensions of size $n$ with the spread of
$$\left.\matrix{
n^k - 1 - n\left({\left({n+1\over 2}\right)^{k-1} - 1}\right),\hfill&
\mbox{ if $n$ is odd,}\hfill\cr\cr
n^k+n-2-\left({n\over2}\right)^{k-1\over2}\left({(n+1)\left({n+2\over2}\right)^{k-1\over2}-2}\right),\hfill&
\mbox{ if $n$ is even, and $k$ is odd,}\hfill\cr\cr
n^k+n-2-\left({n\over2}\right)^{k-2\over2}\left({n+2\over2}\right)\left({n\left({n+2\over 2}\right)^{k-2\over2}-1}\right),\hfill&
\mbox{ if both $n$ and $k$ are even.}\hfill\cr
}\right.
$$
\end{theorem}

\begin{proof}
\begin{figure}
\begin{center}
\epsfysize 2 in \epsfbox{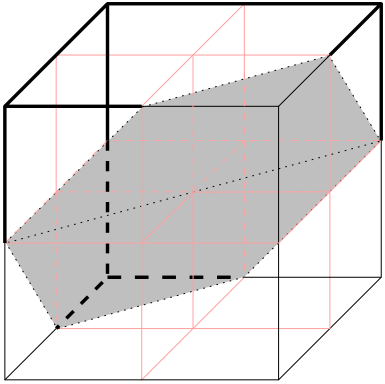}
\caption{Dividing $3$-dimensional matrix into $8$ pieces. Thick lines show
the parts where the border values come from the Herringbone
arrangement for the minima sequence.}
\label{divide}
\end{center}
\end{figure}
We shall assume for simplicity that $n$ is odd. For even $n$ the
argument works in a similar way.  We can divide the matrix into $2^k$
pieces by cutting through the middle of each face, including the
middle line into both sides that are separated by it.  For $k = 3$ see
Figure \ref{divide}. The initial and the last pieces, coordinate-wise,
are entirely within the minima sequence and the maxima sequence
arrangements, respectively.  Now each line lies in two of the $2^k$
pieces. The central lines go through the initial and the last
pieces. The spread in those lines is exactly the maximum difference
between the minima and maxima Herringbone arrangements, as calculated
above. We will show that the spread in any other line does not exceed
that.

There are three types of lines that are not central lines:
\begin{enumerate}
\item[(a)] lines that do not cross the bisecting plane, 
\item[(b)] lines that cross the bisecting plane and the endpoints are 
	neither in the first or the last quadrant of the cube, 
\item[(c)] lines that cross the bisecting plane and one of the endpoints
	is either in the first or the last quadrant of the cube.
\end{enumerate}

If a line does not cross the bisecting plane, then either its minimum
is in the first quadrant or its maximum is in the last quadrant, since
only the first and the last quadrants do not have the bisecting plane
cutting through them. Without loss of generality, let the line lie
completely in the minima arrangement half, and its minimum be in the
first quadrant. Then the line's minimum is at most $\lfloor
n/2\rfloor^k$ away from the parallel central line's minimum, while
it's maximum is at least $\lfloor(n/2)^k\rfloor - \lfloor
n/2\rfloor^k$ away from a central maximum. Thus the spread in the line
is not greater than the spread in a central line.

The minimum in a line of type (b) is not in the first quadrant,
therefore one of the coordinates if the minimum is greater than
$\lceil{n/2}\rceil$. As we have mentioned, Herringbone arrangement is
a fully monotonic arrangement, the values increasing in the direction
of increasing coordinates. Therefore the line's minimum is greater
than the minimum in the parallel central line. Similarly, the line's
maximum is less than the maximum in the parallel central line. Thus
the difference between the line's maximum and minimum, the spread, is
less than that in the parallel central line.

We will now consider a line of type (c). Without loss of generality we
assume that the line's minimum is in the first quadrant. Thus the
maximum is not in the last quadrant, since all the fixed coordinates
of the points on the line are less than $\lceil{n/2}\rceil$. Due to
the full monotonicity of the Herringbone arrangement, both the minimum
and the maximum in the line are less than those in the parallel
central line. We will show that the difference between the central and
the line's minimum is at least the difference between the maxima, thus
making the spread in the line at most that in the central line.
Moreover, we show that this is true for two lines of type (c) that
differ in only one coordinate by $1$, and the spread in the line
closer to the central is at least the spread in the other line. Let
$a_t$ and $a_{t+1}$ be the minima in these lines, and $b_t$, $b_{t+1}$
be the maxima, $t < \lfloor n/2\rfloor$. Let $Corner(k,t)$ be the
number of cells cut off the corner of size $t$ of a $k$-dimensional
cube. Then

$$a_{t+1} - a_t \le (t+1)^k - t^k,$$ and
\begin{eqnarray*}
b_{t+1} - b_t &\ge& (n-t)^k - Corner\left(k, \left(\left\lfloor{n\over2}\right\rfloor-(t+1)\right)k\right) - \left((n-(t+1))^k - Corner\left(k, \left(\left\lfloor{n\over2}\right\rfloor-t)\right)k\right)\right)\\
&=&(n-t)^k - (n-t-1)^k-\left(Corner\left(k, \left(\left\lfloor{n\over2}\right\rfloor-t-1\right)k\right)-Corner\left(k, \left(\left\lfloor{n\over2}\right\rfloor-t\right)k\right)\right)\\
\end{eqnarray*}

Notice that

$$Corner(k,t)=\sum_{i_1=1}^{t}{\sum_{i_2}^{i_1}{\dots\sum_{i_{k-1}}^{i_{k-2}}{i_{k-1}}}} = {t+k-1 \choose k},$$

Therefore

\begin{eqnarray*}
b_{t+1} - b_t &\ge& (n-t)^k - (n-t-1)^k-\left( {\left(\left\lfloor{n\over2}\right\rfloor-t\right)k-k+k-1\choose k} - {\left(\left\lfloor{n\over2}\right\rfloor-t\right)k+k-1 \choose k}\right)\\
&=&(n-t)^k - (n-t-1)^k-\left( {\left(\left\lfloor{n\over2}\right\rfloor-t\right)k-1\choose k} - {\left(\left\lfloor{n\over2}\right\rfloor-t+1\right)k-1 \choose k}\right)\\
\end{eqnarray*}

Thus
\begin{eqnarray*}
(b_{t+1} - b_t)-(a_{t+1} - a_t) &\ge& (n-t)^k - (n-t-1)^k-\left( {\left(\left\lfloor{n\over2}\right\rfloor-t\right)k-1\choose k} - {\left(\left\lfloor{n\over2}\right\rfloor-t+1\right)k-1 \choose k}\right) - ((t+1)^k - t^k)\\
&=& ((n-t)^k - (n-t-1)^k) - ((t+1)^k - t^k) - \left( {\left(\left\lfloor{n\over2}\right\rfloor-t\right)k-1\choose k} - {\left(\left\lfloor{n\over2}\right\rfloor-t+1\right)k-1 \choose k}\right)\\
\end{eqnarray*}

$$ t < \left\lfloor{n\over 2}\right\rfloor, \mbox{ therefore } ((n-t)^k - (n-t-1)^k) - ((t+1)^k - t^k) > 0$$
$${\left(\left\lfloor{n\over2}\right\rfloor-t\right)k-1\choose k} < {\left(\left\lfloor{n\over2}\right\rfloor-t+1\right)k-1 \choose k}, \mbox{ therefore } - {\left(\left\lfloor{n\over2}\right\rfloor-t+1\right)k-1 \choose k} > 0$$

Thus $$(b_{t+1} - b_t)-(a_{t+1} - a_t) \ge 0$$ and the maximum in a
non-central line is further from a central maximum than the minimum in
a non-central line from a central minimum. Therefore the spread in a
line of type (c) is not greater than the spread in a central line.

We have shown that the maximum spread is achieved in the center and is
as calculated above.
\end{proof}

So for a completely filled $k$-dimensional cube the spread is between
the lower bound LB and the upper bound UB, where
$$LB = n^k - 1 - 
\left\lfloor{\left({kn^{k-1}+2\over 2k}\right)^{k\over k-1}}\right\rfloor - 
\left\lfloor{\left({kn^{k-1}\over2k}\right)^{k\over k-1}}\right\rfloor$$
and 
$$UB = \left.\matrix{
n^k - 1 - n\left({\left({n+1\over 2}\right)^{k-1} - 1}\right),\hfill&
\mbox{ if $n$ is odd,}\hfill\cr\cr
n^k+n-2-\left({n\over2}\right)^{k-1\over2}\left({(n+1)\left({n+2\over2}\right)^{k-1\over2}-2}\right),\hfill&
\mbox{ if $n$ is even, and $k$ is odd,}\hfill\cr\cr
n^k+n-2-\left({n\over2}\right)^{k-2\over2}\left({n+2\over2}\right)\left({n\left({n+2\over 2}\right)^{k-2\over2}-1}\right),\hfill&
\mbox{ if both $n$ and $k$ are even.}\hfill\cr
}\right.
$$

This means that in a multiple descriptions system with all channels of
equal capacity the distortion in case of one channel failure is
between LB and UB. We address the case of more than one channel failing
in Section \ref{many_channels}.

\subsection{Completely Filled Cube: Arbitrary Number of Channels Failing}
\label{many_channels}

In the previous section we have obtained the distortion in a multiple
descriptions system with equal capacity channels for the case of one
channel failing. We now consider the possibility of more than one
channel failing. That is, we are concerned with the distortions
$D_{k+1}$ through $D_{2^k-2}$ in the rate-distortion tuples $(\log
n,...,\log n; 0, D_1,...,D_k,D_{k+1},...,D_{2^k-2})$.  In the number
arrangement domain, we are concerned with designing an arrangement
that minimizes the spread in any slice of any dimension.

Notice, however, that the herringbone arrangement maximizes and
minimizes the smalls and bigs sequences, respectively, for any
slice. Thus we can use the same construction
for the algorithm.

The maximum error guaranteed by the algorithm in case of $l$ channel failures
is %(**** - check!!!)
$$
b(\underbrace{\biggl\lceil{n-1\over 2}\biggr\rceil,...,\biggl\lceil{n-1\over 2}\biggr\rceil}_{\lfloor(k-l)/2\rfloor},\underbrace{\biggl\lfloor{n-1\over 2}\biggr\rfloor,...,\biggl\lfloor{n-1\over 2}\biggr\rfloor}_{\lceil(k-l)/2\rceil},\underbrace{*,...,*}_l) -
a(\underbrace{\biggl\lceil{n-1\over 2}\biggr\rceil,...,\biggl\lceil{n-1\over 2}\biggr\rceil}_{\lfloor(k-l)/2\rfloor},\underbrace{\biggl\lfloor{n-1\over 2}\biggr\rfloor,...,\biggl\lfloor{n-1\over 2}\biggr\rfloor}_{\lceil(k-l)/2\rceil},\underbrace{*,...,*}_l),
$$

which if $n$ is odd, equals

$$
n^k-1-\Biggl(\biggl({n+1\over2}\biggr)^{k-l}-1\Biggr){(n+1)^l+(n-1)^l\over2^l},
$$

and if $n$ is even, equals
$$
n^k+n-2-\Biggl(\biggl({n\over2}\biggr)^{\bigl\lceil{k-l\over2}\bigr\rceil}\biggl({n+2\over2}\biggr)^l\biggl(\biggl({n+2\over2}\biggr)^{\bigl\lceil{k-l\over2}\bigr\rceil}-1\biggr)+\biggl({n\over2}\biggr)^l\biggl(\biggl({n\over2}\biggr)^{\bigl\lfloor{k-l\over2}\bigr\rfloor}-1\biggr)+
$$ $$
\biggl({n\over2}\biggr)^{\bigl\lfloor{k+l\over2}\bigr\rfloor}\biggl(\biggl({n+2\over2}\biggr)^{\bigl\lfloor{k-l\over2}\bigr\rfloor}-1\biggr)+\biggl({n-2\over2}\biggr)^l\biggl(\biggl({n\over2}\biggr)^{\bigl\lceil{k-l\over2}\bigr\rceil}-1\biggr)\Biggr).
$$

\subsection{The Infinite Diagonal: $n_i=\infty, 1\le i\le k, m=\infty$}
\label{inf_diag}
We now would like to explore the achievable rate-distortion tuples of
the type $(\log n,\log n,...,\log n,0,D_1,...,D_{2^k-2})$ when the
original information source is quantized into $m < n^k$ integers,
where $\log n$ is channel rate. This means that in the corresponding
matrix the $m$ numbers do not fill the entire matrix. Vaishampayan
\cite{vaish}, \cite{vaish2}, \cite{vaish3} designed a solution for
this case with two channels which is an arrangement of the numbers
into a uniform diagonal. In the next section we examine this
solution. However, to avoid the boundary effects, we will first
consider the ``infinite'' diagonal in this section.  We consider an
arrangement of numbers in a infinite $k$-dimensional diagonal of
thickness $l$, that is, any line in the diagonal has exactly $l$
elements in it. This case is also equivalent to deriving the
achievable rate-distortion tuples for an unbounded discrete
information source and $k$ channels of rate $l$.

\begin{figure}
\begin{center}
\begin{pspicture}(0,0)(3,5)
\rput(-1,2){\begin{tabular}{ccccccccccc}
\cline{4-4}
&&& \multicolumn{1}{|c|}{\nowidth{}} &&&&&&&\\
\cline{5-5}
&&& \multicolumn{1}{|c|}{} & \multicolumn{1}{c|}{} &&&&&&\\
\cline{1-3}\cline{6-6}
\multicolumn{3}{|c|}{} & \multicolumn{1}{|c|}{} & \multicolumn{1}{c|}{} &
    \multicolumn{1}{c|}{} &&&&\\
\cline{1-4}\cline{7-7}
 
& \multicolumn{3}{|@{}l|}{\hbox to 0pt {}} &
    \multicolumn{1}{|c|}{} & \multicolumn{1}{c|}{} &
    \multicolumn{1}{c|}{} &&&&\\
\cline{2-5}\cline{8-8}
 
&& \multicolumn{3}{|c|}{} & \multicolumn{1}{|c|}{} & \multicolumn{1}{c|}{} &
    \multicolumn{1}{c|}{} &&&\\
\cline{3-6}\cline{9-9}
&&& \multicolumn{3}{|c|}{} & \multicolumn{1}{|c|}{} & \multicolumn{1}{c|}{} &
    \multicolumn{1}{c|}{} &&\\
\cline{4-7}
 
&&&& \multicolumn{3}{|c|}{} & \multicolumn{1}{|c|}{} &
     \multicolumn{1}{c|}{} & \multicolumn{1}{c}{} &\\
\cline{5-8}
 
&&&&& \multicolumn{3}{|c|}{} & \multicolumn{1}{|c|}{} &&\\
\cline{6-9}
\end{tabular}
}
\rput(3,3){\epsfxsize 1.5 in \epsfbox{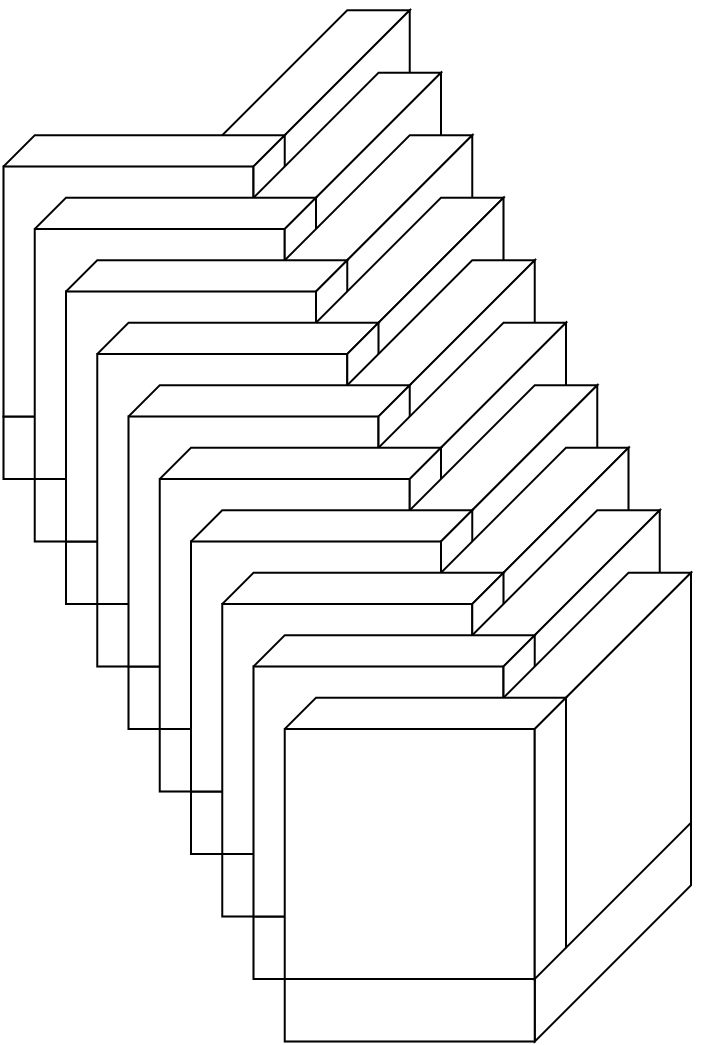}}
\end{pspicture}
\caption{Uniform infinite diagonal herringbone arrangements in two and three dimensions.}
\label{diag_picture}
\end{center}
\end{figure}

Again, we will concentrate on the domain of number arrangement. We
derive the lower bound on the spread in this case in a similar manner
we did in Section \ref{cube} for the case of a complete cube. We use
the herringbone arrangement again to maximize the smalls sequence and
to minimize the bigs sequence. Since the smalls and the bigs sequence
arrangements can start at any point in the diagonal, we will consider
the difference between the sequences relative to the starting points.
The lower bound on the spread is the maximum over all lines of the
difference between the smallest largest number and the largest
smallest nummber.  However, in this case this difference turns out to
be constant.

\begin{definition}
\label{def2}
\end{definition}
A {\em herringbone arrangement} of a $k$-dimensional infinite
$l$-diagonal is defined inductively as follows.  Assign an arbitrary
order to the coordinates of the system $\langle i_1,
i_2,...,i_k\rangle$. A herringbone arrangement of $k$-dimensional
$0$-diagonal is empty. A herringbone arrangement of a
$0$-dimensional diagonal is any arrangement of one number in a cell.

Given a herringbone arrangement of the $k$-dimensional diagonal up to
coordinate $t_i$ in dimension $i$, we define a larger herringbone
arrangement inductively:
\begin{itemize}
\item project the existing arrangement onto the $(k-1)$-dimensional 
hyperplanes $(*,...,*,t_i,*,...,*)$, limited to the diagonal,
\item calculate the volume of each projection,
\item recursively fill the largest volume projection (using coordinate order
to break ties) with the herringbone arrangement for $k-1$ dimensions.
\end{itemize}

We denote the element in the cell $(i_1,...,i_k)$ of the
$k$-dimensional herringbone arrangement by $HB_k(i_1,...,i_k)$.
Examples of a herringbone arrangement of a diagonal are shown in Figure
\ref{diag_picture}.

\begin{lemma}
\label{lemma2}
The herringbone arrangement of values in a $k$-dimensional diagonal
maximizes the smalls sequence -- the ascending list of the smallest
numbers in a line -- for that matrix.
\end{lemma}
The proof is similar to the proof of Lemma \ref{lemma1}.

\begin{corollary}
\label{cor3}
If the smallest number in line $(i_1,...,*,...,i_k)$ of the
herringbone arrangement is $a_j$, then the smallest number in line
$(i_1+1,...,*,...,i_k+1)$ is $$ a_j+\sum_{i=0}^{k-1}{\lfloor
l/2\rfloor^i\lceil l/2\rceil^{k-1-i}}=
\left\{\matrix{
a_j+k(l/2)^{k-1},\hfill			&	\mbox{ if $l$ is even,}\cr\cr
a_j+{(l-1)^k+(l+1)^k\over2^k},\hfill	&	\mbox{ if $l$ is odd.}\cr
}\right.
$$
\end{corollary}

Similarly,
\begin{corollary}
\label{cor4}
If the largest number in line $(i_1,...,*,...,i_k)$ of the herringbone
arrangement is $b_j$, then the largest number in line
$(i_1+1,...,*,...,i_k+1)$ is $$ b_j+\sum_{i=0}^{k-1}{\lfloor
l/2\rfloor^i\lceil l/2\rceil^{k-1-i}}=
\left\{\matrix{
b_j+k(l/2)^{k-1},\hfill			&	\mbox{ if $l$ is even,}\cr\cr
b_j+{(l-1)^k+(l+1)^k\over2^k},\hfill	&	\mbox{ if $l$ is odd.}\cr
}\right.
$$
\end{corollary}

Thus, combining the results of Corollary \ref{cor3} and Corollary
\ref{cor4}, the difference $b_j-a_j$ remains constant along any
diagonal. That is, the difference between the largest and the smallest
numbers in line $(i_1,i_2,...,*,...,i_k)$ equals that of line
$(i_1+s,i_2+s,...,*,...,i_k+s)$ for some integer $s$. To see this,
notice that while we start the smalls arrangement from $0$ at some
point, since the diagonal is infinite, we can continue the arrangement
in the other direction using increasingly smaller numbers. Similarly
with the bigs arrangement, we can continue it in the direction of the
increasing of coordinates using larger numbers. Thus the smalls and
the bigs arrangements are the same arrangements, offset by a certain
value. This arrangements are also ``facing'' opposite directions: the
herringbone arrangement can be viewed as cones stacked into each
other, and in case of the smalls sequence the ``cones'' face the
direction of the coordinate decrease, while in the bigs sequence
``cones'' face the coordinate increase direction. However, the brims
of these cones from both sequences coincide, and since that is where
the smallest and the largest numbers in each line occur, the
difference along any diagonal remains constant.

A consequence of the structure of the herringbone arrangement is the
fact that the difference is maximized over the central diagonal. That
is, if $(t,t,...,t)$ is a cell in the center of the diagonal, then the
maximum difference is achieved for any line $(t+s,t+s,...,*,...,t+s)$,
where $s$ is some integer, and equals to the difference $$HB(t+\lfloor
l/2\rfloor,t,...,t) - HB(t-\lceil l/2\rceil,t,...,t)=$$ $$(\lceil
l/2\rceil-1)\sum_{i=0}^{k-1}{\lfloor l/2\rfloor^i\lceil
l/2\rceil^{k-1-i}}+\lceil l/2\rceil^{k-1}.$$

\subsection{The Incomplete Cube: $n_i=n, 1 \le i \le k; m \le n^k < \infty$}
\label{diag}
Suppose we have an arbitrary quantity of $m \le n^k$ numbers to be
arranged in a $n\times\cdots\times n$ $k$-dimensional cube. This
corresponds to the information source being quantized into $m$
integers, where $\log m$ is less than the combined channel rates.  At
first glance, the diagonal arrangement gives the least
distortion. That is precisely the shape used by Vaishampayan in
\cite{vaish}, \cite{vaish2}, \cite{vaish3}. We show, however, that
diagonal arrangement is not the best possible and a lower distortion
is possible for these rates.

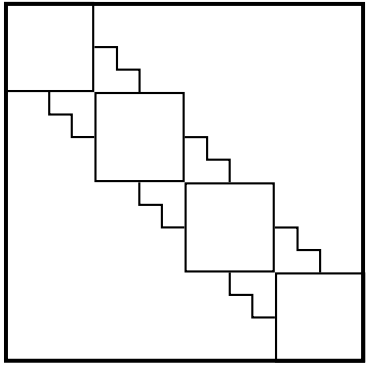
\begin{figure}
\begin{center}
{\psset{unit=.3}
\begin{pspicture}(0,0)(16,16)
\psframe[linewidth=1.5pt](0,0)(16,16)
\multiput(0,16)(4,-4){4}{\psframe(0,0)(4,-4)}
\multiput(2,12)(4,-4){3}{\psline(0,0)(0,-1)(1,-1)(1,-2)(2,-2)}
\multiput(4,14)(4,-4){3}{\psline(0,0)(1,0)(1,-1)(2,-1)(2,-2)}
\end{pspicture}
}
\caption{Possibly an optimal arrangement for for an incompletely filled cube.}
\end{center}
\label{non_diagonal}
\end{figure}

Consider a diagonal arrangement limited to the $n^k$ cube. It is a
restriction of an infinite diagonal, thus the bound on the spread is
the same over the true diagonal part of the arrangement, away from the
boundary effect. However, the boundary parts of the arrangement are
complete cubes of size $\lfloor l/2\rfloor^k$, and the spread there is
the spread in a complete cube derived in Section \ref{cube}. By
comparing the two spreads, in the boundary cubic parts and in the
diagonal part, we can show that the spread on the diagonal is always
greater.  Thus the spread is dominated by the diagonal part, as long
there is a true diagonal part. That is, if $l\le n$, then there is at
least one complete line in each dimension which belongs entirely to
the diagonal, and the spread in this line dominates the overall spread
in the matrix.  But this means that we can increase the size of the
initial cubic part, decrease the width of the diagonal part, thus
decreasing the overall spread.  Not only that, but we can decrease the
spread even more by introducing non-overlapping cubic parts along the
diagonal, thus, by balancing the entire structure, making all of the
cubic part smaller.  This construction i sdemonstrated in Figure
\ref{non_diagonal}. The spread in this arrangement is better than the
diagonal. However currently we do not have a proof whether this
arrangement is optimal or not.

Now, if $l > n$, then the
initial cubic parts of the diagonal overlap. Remembering from Section 
\ref{cube}, the maximum spread of the entirely filled cube is the spread in 
line $(\lceil(n-1)/2\rceil,...,*,...,\lfloor(n-1)/2\rfloor)$, which is
a line in the intersection of the first-quadrant cube and the
last-quadrant cube.  Those are precisely the cubic parts of the
diagonally filled cube.  Since the spread of a cube optimally filled
with $m < n^k$ numbers is at most the spread of the cube filled with
$n^k$ numbers, in the case of the overlapping cubic parts of the
diagonal the spread is the the same as in a completely filled cube.
However, once again, this proves only the upper bound on the optimal
arrangement, not the lower bound.

\subsection{Unequal Channel Rates}
\label{unequal}
There are currently no known existing results for the case of unequal
channel rates, that is, unequal $n_i$'s. Our approach would provide
the first non-trivial lower and upper bounds. Before the final
calculations can be made, however, we need to generalize the concept
of the integer bisecting hyperplane to non-cubic matrices. In 1965
Bresenham \cite{bresenham} gave an algorithm that approximates real
lines on a discrete grid. Since our algorithm for the matrix
arrangement uses two Herringbone arrangements put together at a
bisecting hyperplane, we need to find an appropriate generalization of
Bresenham's algorithm to describe that hyperplane for a matrix with
unequal dimension sizes.

%\subsection{Volume Preserving Transformations}
%It seems that the spread in the matrix for a fixed number of elements
%to be arranged, depends more on the overall size of the matrix (the
%volume, the number of cells in the matrix), rather than on the
%specific size of the individual dimensions. Thus the question arises,
%whether the spread is preserved, or is related, under volume
%preserving transformations.  This question also has a strong practical
%application, since it is related to the problem of choosing fewer
%channels with higher rates versus more channels with lower
%rates when such choice is possible.

%For our algorithm we can compute the answer for the completely filled cube.
%The maximum spread guaranteed by the algorithm is $O(n^k - n(n/2)^{k-1})$.
%Thus given two configurations, one with $t$ channels of capacity $p$ and
%another with $s$ channels of capacity $q$, so that $p^t = q^s$, $p< q$, $t>s$.
%The spread in the first case would be greater than in the second, since
%$2^{t-1} > 2^{s-1}$ and $p^t(1-{1\over 2^{t-1}}) > q^s(1 - {1\over 2^{s-1}})$.
%Our algorithm prefers fewer channels, however, this might not be the case
%for an optimal arrangement.

\section{Conclusions}
We have studied the problem of multiple description scalar
quantizers. We have considered the question of describing the
achievable rate-distortion tuples. The problem has been formulated as
a combinatorial optimization problem of arranging numbers in a
matrix. It has been noted that this formulation is equivalent to a
graph theory problem of finding minimal bandwidth of cartesian
products of cliques.

We have proposed a technique for deriving lower bounds on the
distortion at given channel rates. The approach is constructive thus
allowing an algorithm that gives a fairly tight upper bound. For the
case of two communication channels with equal rates the bounds
coincide thus giving the precise lowest achievable distortion at fixed
rates. To the best of our knowledge, this is the first result
concerning the system with more than two communication channels.

%(*** The result: in case of $l$ channels failing, the exact lower and
%upper bounds)

%\subsection{Open Problems.}
%A quick estimate of the lower bound on the problem can be obtained
%using simple Pigeonhole Principle reasoning \cite{west1}. An
%interesting question therefore to explore is whether the proof can be
%strengthened to give a {\em constructive} proof for a tight lower
%bound. Specifically, whether it can be used to obtain results for the
%general case of unequal channel rates, that is a matrix with
%unequal dimension sizes, and a number of elements to be arranged being
%less than the overall size of the matrix.

%This paper explores the problem of sending a number over multiple channels so
%that a range for the number sent is guaranteed depending on the number of
%successful channels. The number sent generally cannot be recovered even if
%only one channel failed. However, the mechanism of error correcting codes
%seemingly takes an opposite approach of insuring that the number sent can be
%recovered if not too many of its bits were corrupted, regardless on how this
%corruption happened. It would be interesting to explore the possibility of
%combining the two methods.

%This paper explores the problem of sending a number over multiple
%channels so that the maximum distortion is minimized. We have also
%assumed equal channel failure probabilities. Further research should
%take the unequal failure probabilities into consideration and
%concentrate on minimizing the expected distortion.

%In addition, a common generalization is to consider vector, rather
%than scalar, quantizers.

\section{Acknowledgments}
We would like to thank Sergio Servetto for bringing this problem to our 
attention, Douglas West for pointing out the relevance to the graph 
bandwidth problem and helpful suggestions, Ari Trachtenberg, Mitch Harris,
Jeff Erickson, and Ralf Koetter for many fruitful discussions.

\end{document}